\documentclass[aip,pof,unsortedaddress]{revtex4-1}

\usepackage{graphicx}
\usepackage{amsmath}
\usepackage{amssymb}
\usepackage{bm}

\newcommand{\Rey}{\mbox{Re}}
\newcommand{\Kn}{\lambda_{\mbox{s}}/R}
\newcommand{\ls}{\lambda_{\mbox{s}}}
\newcommand{\lv}{\lambda_{\mbox{v}}}
\newcommand{\Ct}{C_{\tau}}
\newcommand{\Cp}{C_{\mbox{\footnotesize p}}}
\newcommand{\Cd}{C_D}
\newcommand{\muR}{\mu_L/\mu_V}

\newcommand{\degrees}{\ensuremath{^\circ}}
\newcommand{\sepang}{\varphi_{\scriptsize\mbox{sep.}}}

\begin{document}


\title{Navier slip model of drag reduction by Leidenfrost vapor layers}


\author{Joseph D. Berry}
\email[]{joe.d.berry@gmail.com}
\affiliation{Department of Chemical and Biomolecular Engineering, University of Melbourne, Parkville, Victoria 3010, Australia}
\author{Ivan U. Vakarelski}
\affiliation{Division of Physical Sciences and Engineering and Clean Combustion Research Centre,
King Abdullah University of Science and Technology (KAUST), Thuwal 23955-6900, Saudi Arabia}

\author{Derek Y. C. Chan}
\affiliation{School of Mathematics and Statistics, University of Melbourne, Parkville, Victoria 3010, Australia}
\affiliation{Department of Mathematics, Swinburne University of Technology, Hawthorn, Victoria 3122, Australia}

\author{Sigurdur T. Thoroddsen}
\affiliation{Division of Physical Sciences and Engineering and Clean Combustion Research Centre,
King Abdullah University of Science and Technology (KAUST), Thuwal 23955-6900, Saudi Arabia}


\date{\today}

\begin{abstract}
Recent experiments found that a hot solid sphere that is able to sustain a stable Leidenfrost vapor layer in a liquid exhibits significant drag reduction during free fall. The variation of the drag coefficient with Reynolds number deviates substantially from the characteristic drag crisis behavior at high Reynolds numbers. Measurements based on liquids of different viscosities show that onset of the drag crisis depends on the viscosity ratio of the vapor to the liquid. Here we attempt to characterize the complexity of the Leidenfrost vapor layer with respect to its variable thickness and possible vapor circulation within, in terms of the Navier slip model that is defined by a slip length. Such a model can facilitate tangential flow and thereby alter the behaviour of the boundary layer. Direct numerical and large eddy simulations of flow past a sphere at moderate to high Reynolds numbers ($10^2 \leq \Rey \leq 4 \times 10^4$) are employed to quantify comparisons with experimental results for the variations of the drag coefficient with Reynolds number and the form of the downstream wake on the sphere. This provides a simple one parameter characterization of the drag reduction phenomenon due to a stable vapor layer that envelops a solid body.
\end{abstract}


\maketitle

\section{Introduction}
The drag on a moving solid body is determined by the nature of fluid flow over its surface and the control of such flow has important consequences on the optimisation of energy use in the design of moving vehicles, ships and aircrafts. Depending upon the speed of the flow and the shape of the object, the flow can separate at a point on the body and result in a pressure drop on the downstream side of the body. For instance, at high Reynolds numbers, the fore-aft difference in pressure distribution on a sphere accounts for around 95\% of the drag force~\cite{Achenbach1972}. The drag force $F_D$ acting on a sphere of radius $R$ moving at velocity $U$ is often characterized in terms of the non-dimensional drag coefficient, $C_D$
\begin{equation}
\label{eq:CD}
\Cd \equiv \frac {F_D}{(\pi R^2) (\frac{1}{2}\rho U^2)}
\end{equation}
where $\rho$ is the fluid density. For a solid sphere that obeys the no-slip boundary condition, $C_D$ is observed to be a universal function of the Reynolds number, $\Rey = 2R\rho U/\mu_L$, where $\mu_L$ is the viscosity of the Newtonian fluid. Since there are no geometrical features such as edges or protrusions to fix the point of flow separation, the location of the separation point on a sphere is extremely sensitive to the local boundary-layer conditions and other surface characteristics. Thus at high Reynolds number, the major influence on the drag coefficient is the position of flow separation on the sphere. 

In the familiar description of the sub-critical behavior of the boundary layer at a solid sphere, the flow near the surface is retarded due to viscous effects. However, as the fluid passes over the front of the sphere, this retardation is counteracted by a negative pressure gradient and the flow remains attached.  As the flow moves over the sphere, the pressure gradient changes sign and acts to oppose motion in conjunction with viscous effects. This causes the fluid velocity in the boundary layer to eventually slow to zero at the stagnation point at which the flow separates from the sphere surface. This gives rise to a region of low pressure in the wake region beyond the stagnation point, resulting in a large pressure difference between the front and the back of the sphere and consequently a large drag force.



\begin{figure}
 \centering
\includegraphics[width=250pt]{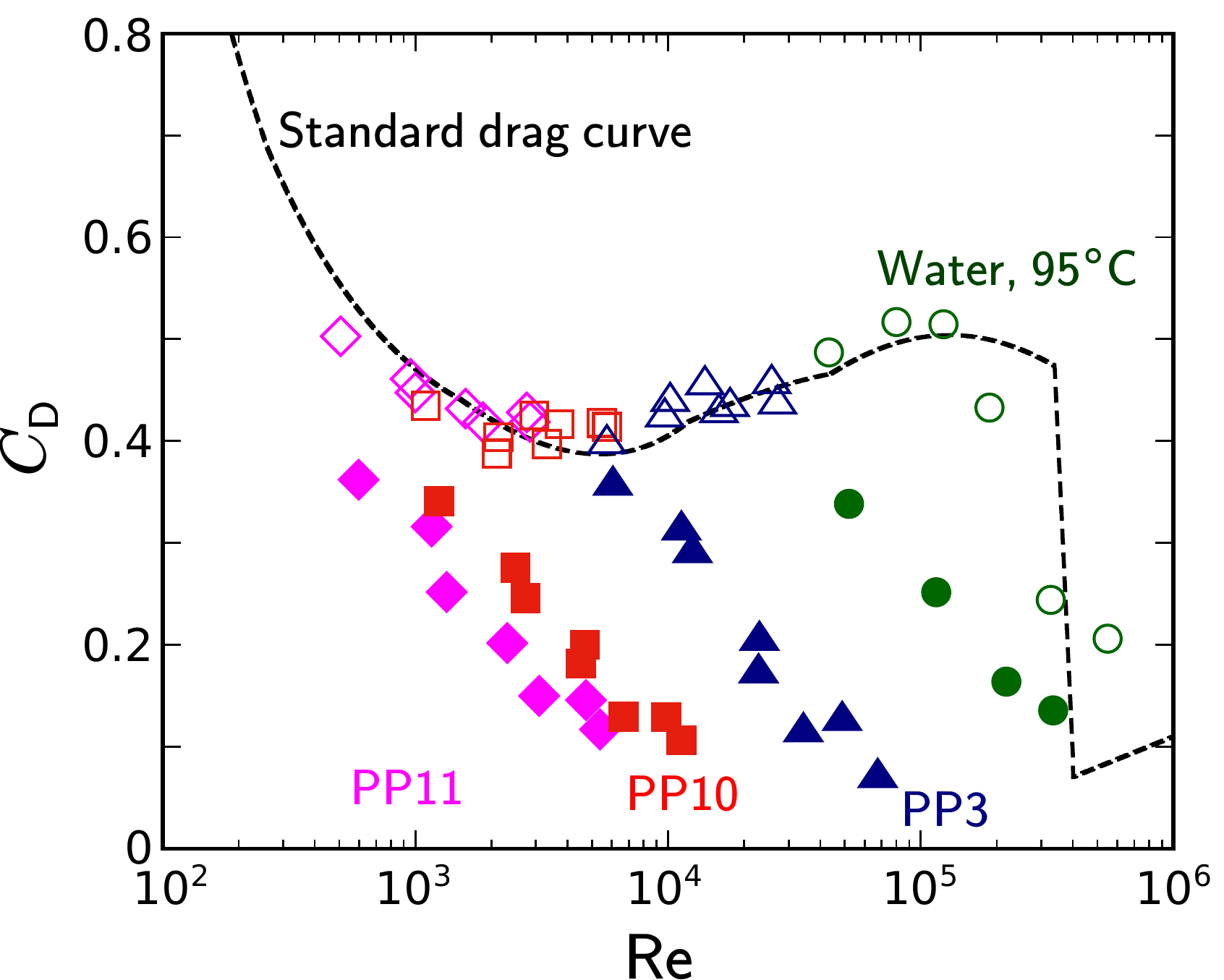}
 \caption{\label{fig:dragCurve} Experimental data showing the variation of the drag coefficient $\Cd$ with Reynolds number, $\Rey$ for hot spheres above the Leidenfrost temperature, see~\citep{Vakarelski2014,Vakarelski2016} for details, free falling in fluorocarbon liquids: PP11 ($\mu_L = 19.2$ mPa s, solid magenta diamonds), PP10 ($\mu_L = 9.6$ mPa s, solid red squares), PP3 ($\mu_L = 1.9$ mPa s, solid blue triangles), and water at 95$\degrees$C ($\mu_L = 0.3$ mPa s, solid green circles). The open symbols represent corresponding results for the $\Cd$ of room temperature spheres without a vapor layer free falling in the same liquids. The viscosity of the vapor is estimated to be $\sim 1.2 \times 10^{-2}$ mPa s for all liquids presented.  The variation of $\Cd$ with $\Rey$ for a no-slip solid sphere in a Newtonian liquid is shown as the dashed curve~\citep{Clift1978,Schlichting1979}.}
 \end{figure}

For a solid sphere characterised by the no-slip boundary condition, the variation of the drag coefficient $\Cd$ with the Reynolds number, $\Rey$ has been studied extensively both experimentally \cite{Tomotika1937,Achenbach1972,Sakamoto1990} and numerically \cite{Mittal1999,Tomboulides2000,Constantinescu2003,Constantinescu2004,Yun2006,hoffman2006,Jones2008}. At low $\Rey \sim 0$, the flow around the sphere is axisymmetric, steady and fully attached, and the drag coefficient varies as $\Cd = 24/\Rey$, with the no-slip or zero tangential velocity boundary condition on the sphere surface. Flow separation occurs at $\Rey \approx 5$ and the axisymmetry of the wake is broken at $\Rey \approx 210$. At $\Rey \approx 270$, the wake becomes unsteady and planar asymmetry such as vortex shedding begins to occur. Above $\Rey \approx 375$ the planar symmetry is broken and the wake becomes both unsteady and asymmetric thereafter.  For Reynolds numbers in the range $10^3 \leq \Rey \leq 4 \times 10^5$, the drag coefficient of a no-slip sphere is relatively independent of Reynolds number, $\Cd \sim 0.4$. As the Reynolds number increases further, beyond about $\Rey \sim 5 \times 10^5$, the drag coefficient on the no-slip sphere undergoes a sharp drop to $\Cd \sim 0.1$ as the boundary layer transitions to turbulence. This phenomenon is the well-known ``drag crisis'' (see Figure~\ref{fig:dragCurve}).


In contrast to a no-slip sphere, analytical studies of flow around a sphere with the free-slip or zero tangential stress boundary condition show that the drag coefficient follows the Hadamard-Rybczynski result, $\Cd = 16/\Rey$, at $\Rey \sim 0$ \cite{Hadamard11, Rybczynski11}. With increasing $\Rey$, the wake remains axisymmetric and steady for low, moderate and high Reynolds numbers \cite{Moore1963}, and no separation is predicted to occur. For Re $\gg 1$, the wake thickness varies as $O(\Rey^{-1/4})$ as the drag coefficient assumes the asymptotic form~\cite{Moore1963} 
\begin{equation}
\label{eq:Moore}
\Cd \approx \frac{48}{\Rey}\left(1-\frac{2.2}{\sqrt{\Rey}}\right).
\end{equation}
Thus, the drag coefficient of a free-slip sphere decreases monotonically for large Reynolds numbers and the flow remains fully attached. However, such limiting behavior has yet to be observed because a sphere with a free-slip or zero tangential stress surface has yet to be realised.
Nevertheless, it is pertinent to note that at $\Rey \sim 0$, the drag coefficient, $\Cd$ only changes by a factor $2/3$ between the no-slip and the free-slip boundary condition and Equation \ref{eq:Moore} provides a point of reference as to the limiting behaviour of a free-slip body in the limit Re $\gg 1$.

Recent experimental studies using solid spheres have demonstrated the possibility of using a surface bound vapor layer that is maintained by the Leidenfrost effect generated by a hot surface held at a temperature well above the boiling point of the liquid \cite{Shirtcliffe2009,McHale2009,McHale2010,McHale2011,Vakarelski2011, Vakarelski2012,Vakarelski2013,Vakarelski2014} or using a thin surface mass transfer layer maintained by a melting solid surface \cite{Vakarelski2015} to move the point at which flow separates towards the rear of the sphere and thereby achieve a corresponding reduction in the drag.  The thickness of these surface layers are of order hundreds of micrometers, extremely small relative to the centimeter radius of the sphere.  The early studies on drag reduction caused by the presence of a stable Leidenfrost vapor layer on a sphere found significant reduction only at high Reynolds numbers ($\Rey \gtrsim 2 \times 10^4$)~\cite{Vakarelski2011, Vakarelski2016}. However, more recent experiments showed that the Reynolds number at the onset of significant drag reduction is dependent on the viscosity of the gas in the vapor layer, $\mu_V$ relative to the viscosity of the surrounding liquid, $\mu_L$ \citep{Vakarelski2016}.  Indeed, significant drag reduction was observed for large values of the viscosity ratio, $\muR \sim 1900$ at $\Rey \sim 10^3$, that is well below the critical $\Rey$ value, $\Rey \sim 5 \times 10^5$, that marks the transition to turbulence for solid spheres without surface vapor layers (see Figure 1).  Results from such experimental observations are summarized in Figure~\ref{fig:dragCurve} for hot spheres with sustained Leidenfrost vapor layers undergoing free fall in four fluorocarbon liquids that span a 10-fold variation in viscosities \citep{Vakarelski2014,Vakarelski2016}. Collapse of the drag coefficient data onto a single curve was achieved when plotted as a function of the parameter $(\muR)\Rey$, though this is not necessarily an indication of a universal master curve \citep{Vakarelski2016}. 

Attempts at modeling of the effective boundary condition that would describe laminar fluid flow over solid surface covered by a thin vapor layer have been reviewed recently~\cite{McHale2017}. Such models specify the constant thickness of the vapor layer, the viscosity and density of the vapor phase and allow for the possibility of circulation of the vapor phase within the thin layer. The objective is to obtain a relation between the vapor layer properties to a slip length of the Navier model. The results have then been used to interpret numerical solutions of the Navier-Stokes equation for the same model of a concentric vapor layer of constant thickness around a sphere up to $\Rey \sim 100$~\cite{McHale2011,Gruncell2013}. These models assume no mass transfer between the vapor layer and the surrounding liquid. However, for free falling spheres that are covered by a sustained Leidenfrost vapor layer as in our experiments, vapor is continually generated by the hot surface of the sphere and then shed into the wake at the rear. Therefore these explicit models will require more parameters such as vapor density and viscosity ,in addition to the assumption of constant vapor layer thickness, in order to describe the key features of the flow. For a centimeter diameter sphere, the vapor layer thickness can only be estimated to be in the range of $50 - 200$ $\mu$m, thus making precise measurement and modeling of the vapor layer properties problematic.


To circumvent the above practical limitations that preclude specification of the detailed features of the surface vapor layer, we study the predictions of the full Navier-Stokes equations with a Navier slip boundary condition and compare them to experimental observations. Preliminary results suggest that this simplified model was able to capture the drag reduction observed experimentally~\citep{Vakarelski2016}.  The Navier slip boundary condition has often been used to characterise the flow of a liquid over a thin layer of gas next to a wall \cite{Lauga2003,Lauga2007}. When applied to Leidenfrost scenarios, the Navier slip model has been previously used to derive the variation of the Navier slip length $\ls$ with the vapor layer thickness, $\lambda_V$ and the viscosities of the vapor, $\mu_V$ and liquid, $\mu_L$ \cite{Vinogradova1995, Tretheway2004,McHale2011,Busse2013,Saranadhi2016,McHale2017}. These results, obtained in the limit $\Rey = 0$ suggests the relationship 
\begin{equation}
\label{eq:slipToViscRatio}
\ls \sim \left(\frac{\mu_L}{\mu_V}\right) \lv.
\end{equation}
As yet, this relationship has not been tested at moderate to high $\Rey$ flows up to the drag crisis.

In the context of flow over a sphere, the Navier slip model has the advantage of direct, unambiguous calculation of physical quantities such as the drag force and the wake separation angle because there is no longer separate vapor and liquid regions as in the model of \citet{Gruncell2013}. Further, the Navier slip model is characterized by only one parameter, the Navier slip length, $\ls$. As $\ls$ is increased from zero, the separation angle will move towards the rear of the sphere, until the free-slip limit of Equation \ref{eq:Moore} is reached when the flow remains fully attached. In effect, the parameter $\ls$ allows us to quantify the variation of drag coefficient $\Cd$ with separation angle $\sepang$. 
In this paper we use the Navier slip model to capture the effects of the Leidenfrost vapor layer on the drag force and wake shape over the experimental range of Reynolds numbers, $10^2 \leq \Rey \leq 4 \times 10^4$~\cite{Vakarelski2016}. We compare our results to existing experimental data in order to quantify the relationship between the Navier slip length, $\ls$ and measurable quantities such as the Leidenfrost vapor layer thickness, $\lv$ and the viscosities $\mu_V$ and $\mu_L$ of the vapor and the liquid respectively.

\section{Method}
\begin{figure}
 \centering
\includegraphics[width=180pt]{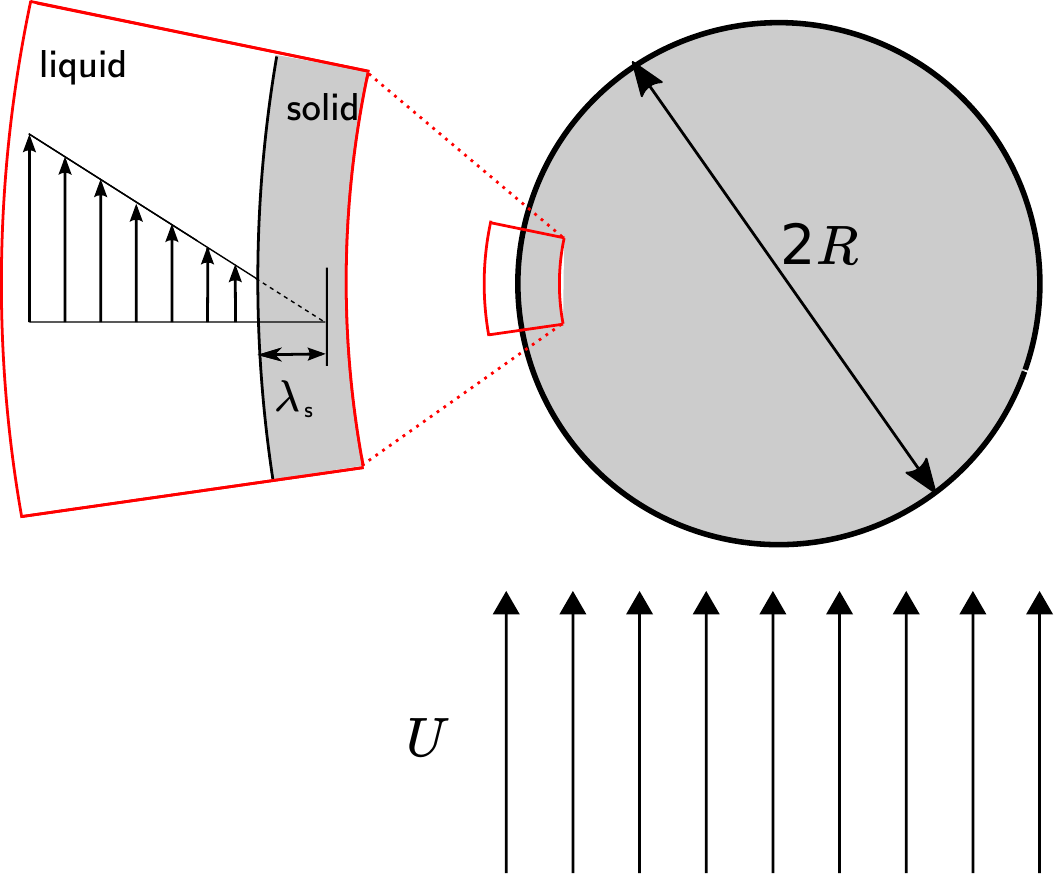}
 \caption{\label{fig:schematic}Schematic of a partial-slip sphere with slip length $\ls$ in a uniform flow of velocity $U$.}
 \end{figure}

 For the Leidenfrost sphere in free-fall experiments, vapor is continually created at the surface of the super-heated sphere and is subsequently swept downstream along the sphere and into the wake.
In this study we do not attempt to capture the dynamics inside the vapor layer. Instead, as in \citet{Vakarelski2016} and in low Reynolds number models as in \citet{Gruncell2013} , we assume that the vapor layer has a constant thickness, $\lv$ that is much smaller than the sphere radius ($\lv \ll R$), and thus affects the flow through a modification of the usual no-slip boundary condition at the surface of the sphere (Figure \ref{fig:schematic}). In this simple model of the Leidenfrost vapor layer, we assume that the flow around the sphere is isothermal, and that the vapor layer thickness is constant and uniform.  These assumptions are then represented by the Navier slip boundary condition in non-dimensional form~\cite{Lamb1932,Lauga2003,Lauga2007,Swan2008} 
\begin{equation}
\label{eq:slipND}
\bm{t}^{(i)} \cdot \bm{u} = \frac{1}{2}\frac{\lambda_{\mbox{s}}}{R} \; \bm{t}^{(i)}\bm{n}:\bm{\tau}.
\end{equation}
Here $\ls/R$ is the constant slip length divided by the sphere radius, $\bm{t}^{(i)}$ and $\bm{n}$ are the unit vectors tangential and normal to the surface respectively, $\bm{u}$ is the fluid velocity and $\tau$ is the fluid shear stress. 
Equation \ref{eq:slipND} has been non-dimensionalised with velocity scale $U$, length scale $2R$, and stress scale $\mu_L U/R$.
 
 
To model the terminal velocity state of a sphere in free fall in an incompressible Newtonian liquid, we use direct numerical simulations (DNS) for $\Rey \le 10^4$, and large eddy simulations (LES) for $\Rey \geq 10^4 $ with a dynamic Smagorinsky turbulence model that is consistent with previous studies  \cite{Germano1993,Constantinescu2003,Constantinescu2004, Vakarelski2016}. Simulations were undertaken on a rectangular domain extending $32R$ upstream and $42R$ downstream of the sphere centre, and $32R$ in the directions normal to the flow.  The normal velocity at the upstream boundary was specified as a constant velocity $U$, with $U $ chosen to give the desired Reynolds number. The corresponding tangential velocities were set to zero. The tangential velocity in the flow direction on the four boundaries normal to the flow were also specified as $U$, with the other two velocity components set to zero. The downstream boundary was specified as an outlet, with zero normal velocity gradient.

The first mesh point normal to the sphere surface was located within one dimensionless viscous unit:
\begin{equation}
\Delta r^+=\frac{\rho u_{\tau}r}{\mu} = 1
\end{equation}
Here $r$ is the distance from the sphere surface, $u_{\tau}=\sqrt{\tau_w/\rho}$ is the (maximum) friction velocity,  and $\tau_w$  is the (maximum) surface shear stress. Consistent with previous numerical simulations \citep{Constantinescu2004, Vakarelski2016}, the friction velocity was estimated as $0.04U$ \emph{a priori},  and then checked for validity \emph{a posteriori}. A minimum of 7 mesh points were positioned within 10 wall units of the sphere, and the maximum size of elements on the sphere surface was approximately 5 - 30 wall units, depending upon the size of Reynolds number chosen. The resultant mesh size was approximately 6.26 million elements.
 
\section{Results and Discussion}
\begin{figure}
\centering
 \includegraphics[width=0.6\textwidth]{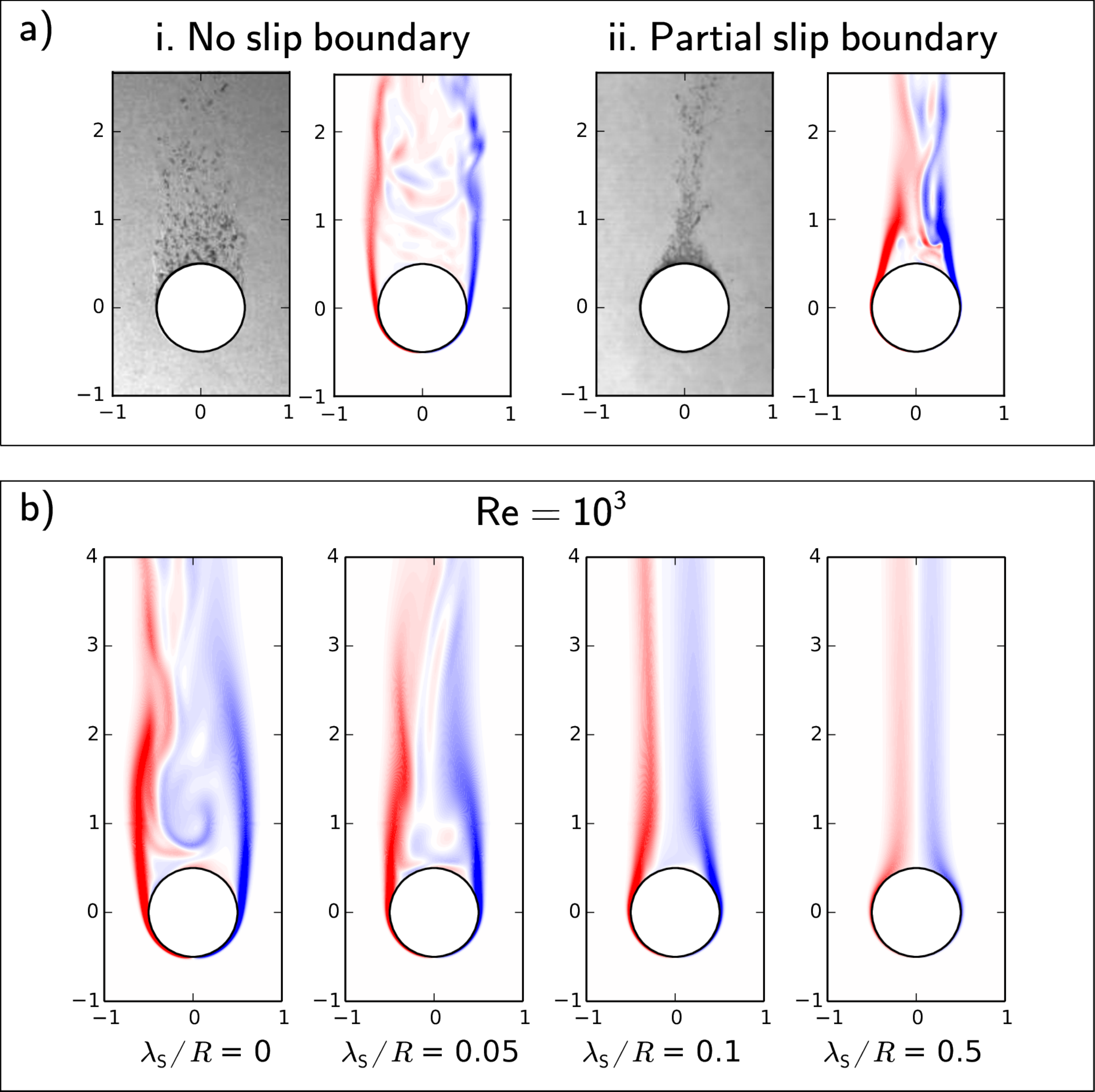}
\caption{\label{fig:Contours} a) Comparisons of instantaneous wakes between experiment (performed in PP11) and simulation based on the Navier slip model with the indicated slip lengths, $\ls$: i) No vapor layer: $\Rey = 2.3 \times 10^3$ - experiment (left) and simulation with $\ls/R = 0$ (right), ii) With Leidenfrost vapor layer: $\Rey =  5.8 \times 10^3$ - experiment (left) and simulation with $\ls/R = 0.045$ (right). The simulation results show contours of instantaneous out-of-plane vorticity. b) Contours of instantaneous out-of-plane vorticity for $\Rey = 10^3$ and four different values of slip length $\ls/R$.}
 \end{figure}

In Figure \ref{fig:Contours}a we show examples of the instantaneous experimental wake patterns on spheres without and with a Leidenfrost vapor layer falling in the perfluorocarbon liquid PP11. Also shown are the numerical results for a no-slip sphere and a sphere with the Navier slip boundary condition at similar Reynolds numbers.  
It is clear that the Navier slip model is able to reproduce the point of separation of the boundary layer, and the subsequent wake pattern, observed experimentally for spheres encased by thin vapor layers.

In Figure \ref{fig:Contours}b we show the effect of slip length on the wake at a fixed Reynolds number $\Rey = 10^3$. For the no-slip case ($\Kn = 0$), the flow is unsteady and asymmetric. As the slip length increases, the flow separation point moves downstream along the sphere towards the rear stagnation point. For $\Kn \gtrsim 0.1$, the wake becomes steady, and as the slip length increases further, the flow becomes axisymmetric and remains fully attached. 

\begin{figure*}
 \includegraphics[width=0.9\textwidth]{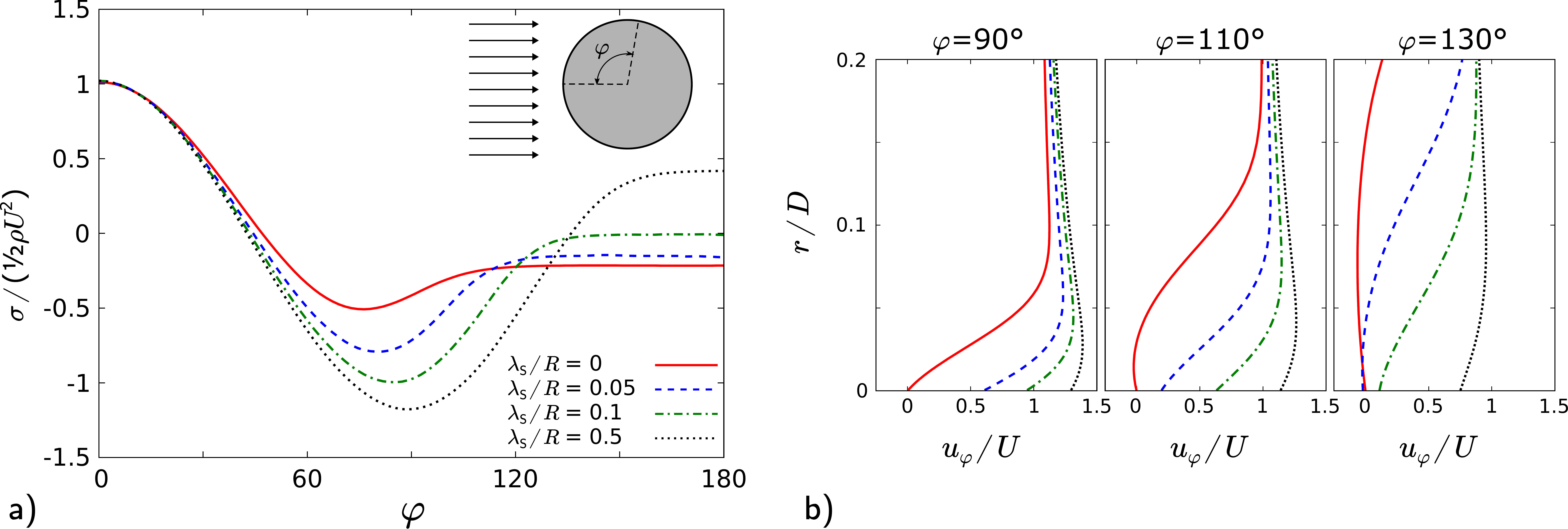}
  \caption{\label{fig:Re1e3} Flow past a sphere at Reynolds number $\Rey = 10^3$. a)  Distributions of mean normal stress $2\sigma/\rho U^2$ on the sphere surface, and b) tangential velocity profiles near the sphere at three positions for slip lengths $\Kn~=~0,\;0.05,\;0.1\; \&\;0.5$. Here $r$ is the coordinate in the normal direction ($r = 0$ on the sphere surface) and $u_\varphi$ is the velocity in the tangential direction.}
 \end{figure*}

In Figure \ref{fig:Re1e3} we show the normal mean stress distributions and tangential velocity profiles for flow past a sphere at $\Rey = 10^3$ for various values of dimensionless slip length, $\Kn$. 
 For a no-slip sphere without a vapor layer ($\Kn = 0$) in incompressible flow, there is no normal component of the viscous stress tensor at the sphere surface because mass conservation stipulates that $\partial u_r/\partial r = 0$, and the only normal stress acting on the sphere is the pressure. However, for non-zero slip lengths, a normal component of viscous stress can exist at the sphere surface because the surface tangential velocities and their corresponding tangential derivatives, at the surface are in general non-zero. This then gives rise to a finite value of $\partial u_r/\partial r$ and a normal stress at the surface.  
 Finite values of $\Kn$ have a marked effect on the normal stress distribution, with separation delayed considerably in comparison to the non-slip case ($\Kn = 0$), resulting in a much narrower wake. 
 When $\Kn \neq 0$, the viscous retardation of the fluid in the boundary layer is not as strong because the fluid is able to move along the surface of the sphere (Figure \ref{fig:Re1e3}b). This means that the flow remains attached beyond the expected separation point for a no-slip sphere before the adverse pressure gradient is able to slow the fluid down enough to cause separation. 
 The resulting delay in separation decreases the size of the wake region, resulting in a smaller region of low pressure on the surface of the sphere and also  a larger back pressure (Figure \ref{fig:Re1e3}b). These two effects both act to decrease the total drag force exerted on the sphere.

\begin{figure}
\includegraphics[width=0.9\textwidth]{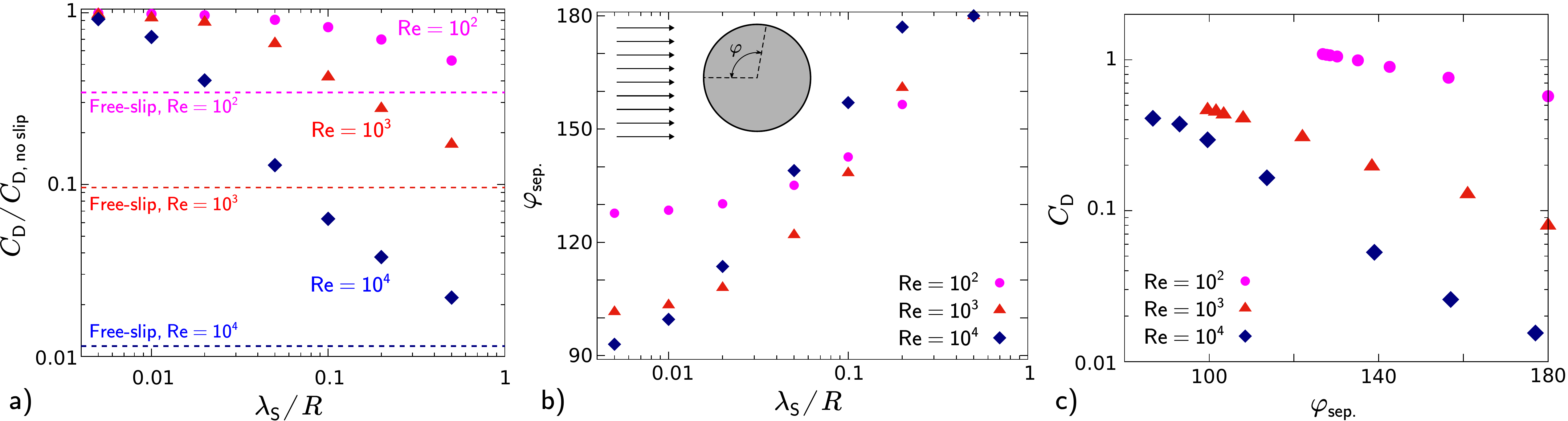}
 \caption{\label{fig:DragAngle} a) Drag coefficient normalised by the drag coefficient for a no-slip sphere, b) separation angle as a function of dimensionless slip length $\Kn$, and c) drag coefficient as a function of separation angle,  for Reynolds numbers $\Rey = 10^2$, $10^3$ and $10^4$. The dashed lines in a) indicate the normalised drag coefficients for a free-slip sphere at each Reynolds number, calculated with the aid of Equation \ref{eq:Moore}.}  
 \end{figure}

This reduction in drag force is demonstrated clearly in Figure \ref{fig:DragAngle}, where we plot the drag coefficient, $C_D$ and separation angle, $\sepang$ for three Reynolds numbers: $\Rey = 10^2, 10^3$ and $10^4$, over a wide range of slip lengths $\Kn$.
 Here the drag coefficients have been normalised by the drag coefficient of a no-slip sphere at the same Reynolds number. The decrease in drag coefficient with increasing slip length corresponds with a delay in the angle at which the flow separates from the sphere (Figure \ref{fig:DragAngle}b).  For $\Kn > 0.5$, the flow remains fully attached to the sphere, $\sepang \approx 180^{\circ}$, and the drag coefficient approaches the predicted value for a free-slip sphere, given by Moore's formula (Equation \ref{eq:Moore}). The results demonstrate that the effect of the slip length becomes more marked as the Reynolds number increases.  For $\Rey = 10^4$, slip lengths $\Kn \gtrsim 0.02$, have significant effect on the separation angle and subsequent drag coefficient. In contrast, slip lengths $\Kn \lesssim 0.1$ have little effect at $\Rey = 10^2$. The effect of separation angle on the drag coefficient is shown in Figure \ref{fig:DragAngle}c. It is clear that a small delay in separation angle has a profound effect on the drag coefficient, with an exponential decay in drag coefficient as the separation angle moves from $\sim 90\degrees$ for the no-slip case to $180\degrees$ for the free-slip case. This explains the increased sensitivity with Reynolds number to the presence of a slip length at the surface of the sphere, because the drag coefficient for the fully attached, free-slip case is proportional to $\Rey^{-1}$ (Equation \ref{eq:Moore}), whereas the drag coefficient for the no slip case is relatively independent of Reynolds number ($\Cd \sim 0.4$). Thus, as the Reynolds number increases for fixed slip length $\ls$, the drag coefficient will also decrease.

\begin{figure}[]
\includegraphics[width=0.99\textwidth]{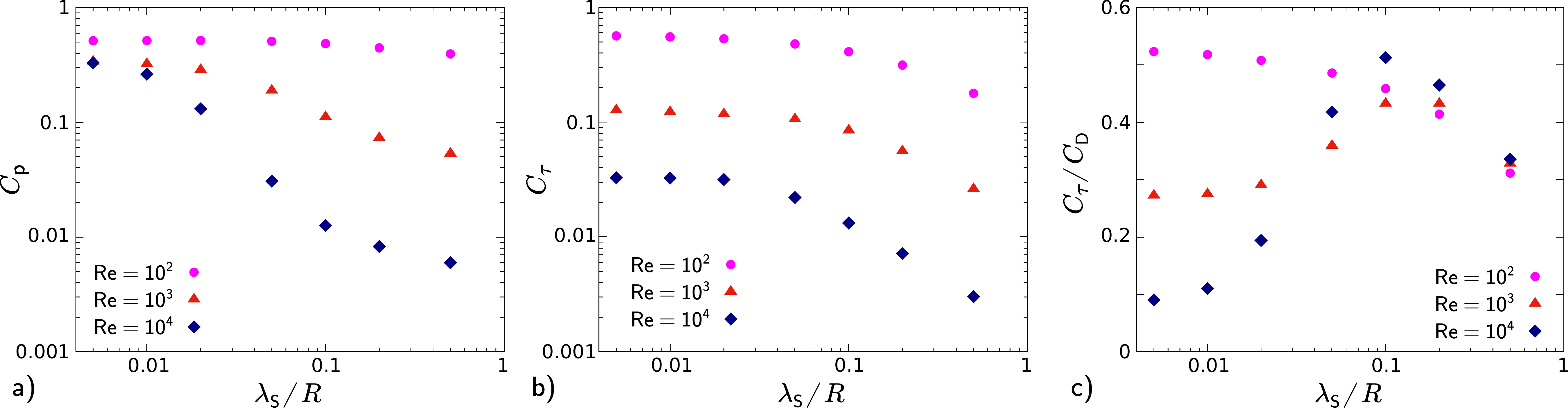}
 \caption{\label{fig:SkinFriction} a) Pressure drag coefficient, $\Cp$, b) skin friction coefficient, $\Ct$, and c) ratio of skin friction drag coefficient to total drag coefficient, $\Ct/\Cd$, as functions of dimensionless slip length, $\Kn$, for Reynolds numbers $\Rey = 10^2$, $10^3$ and $10^4$. }
 \end{figure}

The delay in flow separation decreases the overall contribution of pressure drag, but may lead to an increase in skin friction drag because the flow remains attached over a greater portion of the sphere. To examine if this is the case, in  Figure \ref{fig:SkinFriction} we plot the individual contributions to the drag coefficient: $\Cd \equiv \Cp + \Ct$, from the normal stress, $\Cp$ and from the skin friction, $\Ct$ to the overall drag coefficient, along with the ratio $C_\tau/\Cd$.
It is clear that for $\Rey > 10^2$ the delay in separation angle induced even by small slip lengths has little effect on the skin friction coefficient $\Ct$, but has a significant effect on the pressure coefficient $\Cp$. The pressure drag coefficient decreases markedly due to the delay in separation induced by finite slip lengths, with little effect on the skin friction coefficient, and consequently the total drag coefficient decreases commensurately.
At higher slip lengths the skin friction coefficient decreases monotonically towards the free-slip limit of zero, and the drag coefficient consists entirely of drag due to pressure.

\begin{figure}
\centering
 \includegraphics[width=0.9\textwidth]{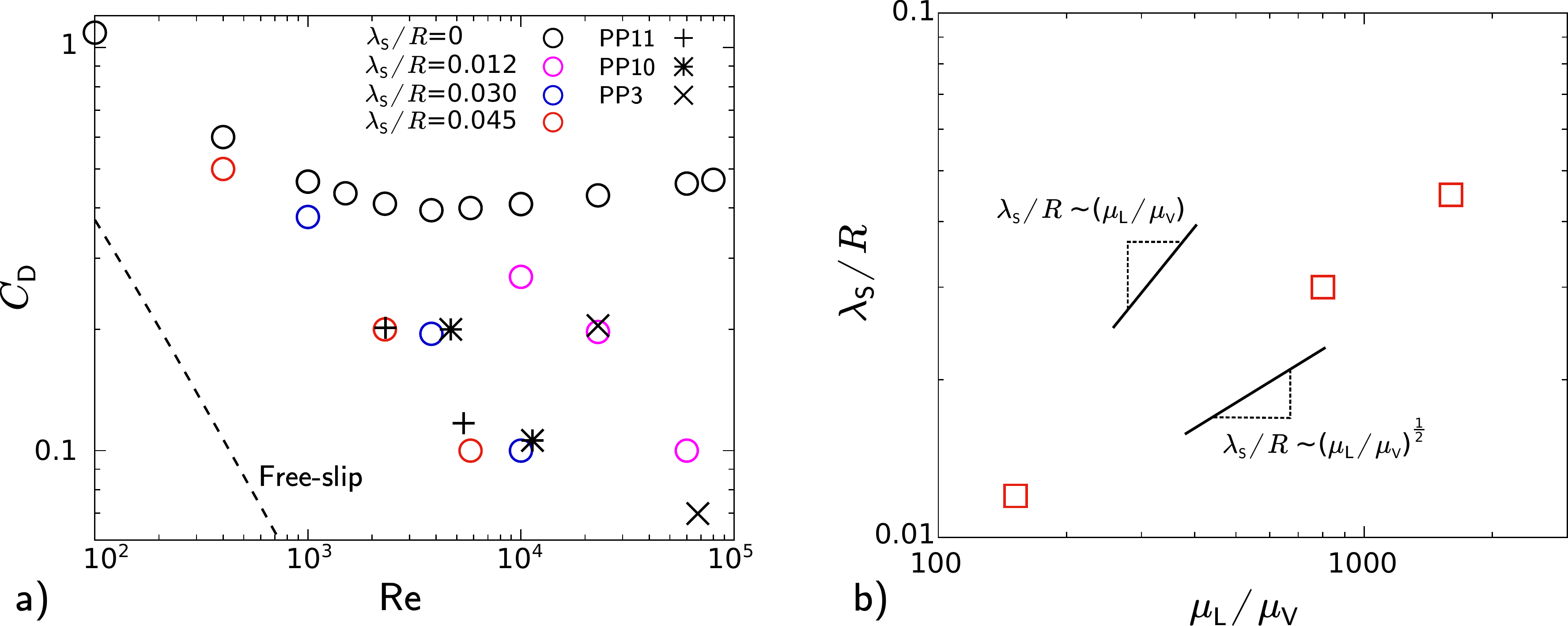}
  \caption{\label{fig:ExpComp} a) Variation of drag coefficient with Reynolds number for three values of $\Kn$. The experimental results of \citet{Vakarelski2016} are also depicted for a sphere radius of $R = 20$ mm. b) Numerical slip lengths matching the experimental data for three viscosity ratios.}
 \end{figure}

In Figure \ref{fig:ExpComp}a, we compare the simulation results to the experimental results of \citet{Vakarelski2016} for spheres of fixed radius $R = 20$ mm. The experimental results depicted include the results for spheres sustaining vapor layers in three perfluorocarbon liquids PP3, PP10, and PP11, of viscosity ratios $\muR \sim 150$, 800 and 1600 respectively. In these experiments the vapor layer thickness was estimated to be $150 \pm 50$ $\mu$m for all cases.
The results of the simulation show excellent agreement with experimental observations, demonstrating that a partial slip boundary condition is sufficient to capture the effect of the presence of a vapor layer on the suface of a solid sphere. In Figure \ref{fig:ExpComp}b, we plot the numerical slip length that best matches the experimental datasets against the corresponding experimental viscosity ratio $\muR$.
Equation \ref{eq:slipToViscRatio} implies that the slip length $\ls$ varies linearly with the viscosity ratio for low $\Rey$ flows. It is clear from Figure \ref{fig:ExpComp}b that assuming this relationship for moderate to high Reynolds number flows will over-predict this dependence, providing strong evidence that the variation of slip length with viscosity ratio is not universal for all Reynolds numbers.


\section{Conclusion}
We have shown that the significant drag reduction exhibited by hot spheres that are capable of sustaining a stable Leidenfrost vapor layer on its surface can be modeled numerically using a Navier slip boundary condition, characterised by the slip length $\ls$.  As the slip length decreases from high to very low for fixed Reynolds number, the flow past the sphere transitions from steady attached flow to separated flow to vortex shedding arising from complex unsteady behaviour. The presence of a finite tangential velocity on the surface of the sphere enables the flow to resist the adverse pressure gradient for longer, delaying flow separation and leading to a smaller low pressure region behind the sphere and a larger back pressure. As a direct consequence, the magnitude of pressure drag acting on the sphere decreases with increasing slip length. 

As the Reynold number is increased, small slip lengths have a profound effect on the flow and resultant drag reduction, due to the delay in flow separation. The increased sensitivity to Reynolds number is due to the dependence of the drag coefficient for a free-slip sphere on the Reynolds number ($\Cd \sim \Rey^{-1}$), whereas the drag coefficient for a no-slip sphere is relatively independent of $\Rey$. Thus a small delay in separation angle due to a finite slip length leads to a large decrease in drag coefficient at high $\Rey$.
Analysis of low $\Rey$ flows in the literature suggests that the slip length is of magnitude $\ls \sim (\mu_L/\mu_V)\lv$ \cite{Vinogradova1995, Tretheway2004,McHale2011,Busse2013,Saranadhi2016,McHale2017}. Through comparison to experimental results in the literature, we have demonstrated that, at moderate to high Reynolds numbers, the slip length $\ls$ is a function of the viscosity ratio $\mu_L/\mu_V$, but does not follow the form suggested by low $\Rey$ flow analysis.

To enable simulations to model higher Reynolds numbers near and above transition in the boundary layer ($\Rey \gtrsim 10^5$) the turbulence model needs to be formulated for a partial-slip surface. For the Reynolds numbers considered here, the flow becomes turbulent in the wake, downstream of the sphere surface, and standard turbulence models are applicable. For $\Rey \gtrsim 4 \times 10^5$, the boundary layer becomes turbulent and the standard turbulence model breaks down. More sophisticated models that account for the correct asymptotic behaviour near a gas/liquid interface are available \cite{Liovic2007}, but significant modification is required in order to model this system at and above transition.


\begin{thebibliography}{37}%
\makeatletter
\providecommand \@ifxundefined [1]{%
 \@ifx{#1\undefined}
}%
\providecommand \@ifnum [1]{%
 \ifnum #1\expandafter \@firstoftwo
 \else \expandafter \@secondoftwo
 \fi
}%
\providecommand \@ifx [1]{%
 \ifx #1\expandafter \@firstoftwo
 \else \expandafter \@secondoftwo
 \fi
}%
\providecommand \natexlab [1]{#1}%
\providecommand \enquote  [1]{``#1''}%
\providecommand \bibnamefont  [1]{#1}%
\providecommand \bibfnamefont [1]{#1}%
\providecommand \citenamefont [1]{#1}%
\providecommand \href@noop [0]{\@secondoftwo}%
\providecommand \href [0]{\begingroup \@sanitize@url \@href}%
\providecommand \@href[1]{\@@startlink{#1}\@@href}%
\providecommand \@@href[1]{\endgroup#1\@@endlink}%
\providecommand \@sanitize@url [0]{\catcode `\\12\catcode `\$12\catcode
  `\&12\catcode `\#12\catcode `\^12\catcode `\_12\catcode `\%12\relax}%
\providecommand \@@startlink[1]{}%
\providecommand \@@endlink[0]{}%
\providecommand \url  [0]{\begingroup\@sanitize@url \@url }%
\providecommand \@url [1]{\endgroup\@href {#1}{\urlprefix }}%
\providecommand \urlprefix  [0]{URL }%
\providecommand \Eprint [0]{\href }%
\providecommand \doibase [0]{http://dx.doi.org/}%
\providecommand \selectlanguage [0]{\@gobble}%
\providecommand \bibinfo  [0]{\@secondoftwo}%
\providecommand \bibfield  [0]{\@secondoftwo}%
\providecommand \translation [1]{[#1]}%
\providecommand \BibitemOpen [0]{}%
\providecommand \bibitemStop [0]{}%
\providecommand \bibitemNoStop [0]{.\EOS\space}%
\providecommand \EOS [0]{\spacefactor3000\relax}%
\providecommand \BibitemShut  [1]{\csname bibitem#1\endcsname}%
\let\auto@bib@innerbib\@empty
\bibitem [{\citenamefont {Achenbach}(1972)}]{Achenbach1972}%
  \BibitemOpen
  \bibfield  {author} {\bibinfo {author} {\bibfnamefont {E.}~\bibnamefont
  {Achenbach}},\ }\href {\doibase 10.1017/S0022112072000874} {\bibfield
  {journal} {\bibinfo  {journal} {Journal of Fluid Mechanics}\ }\textbf
  {\bibinfo {volume} {54}},\ \bibinfo {pages} {565} (\bibinfo {year}
  {1972})}\BibitemShut {NoStop}%
\bibitem [{\citenamefont {Tomotika}(1937)}]{Tomotika1937}%
  \BibitemOpen
  \bibfield  {author} {\bibinfo {author} {\bibfnamefont {S.}~\bibnamefont
  {Tomotika}},\ }\href@noop {} {\emph {\bibinfo {title} {{Laminar Boundary
  Layer on the Surface of a Sphere in a Uniform Stream}}}},\ \bibinfo {type}
  {Tech. Rep.}\ (\bibinfo  {institution} {Aeronautical Research Committee},\
  \bibinfo {year} {1937})\BibitemShut {NoStop}%
\bibitem [{\citenamefont {{Sakamoto, H. Haniu}}(1990)}]{Sakamoto1990}%
  \BibitemOpen
  \bibfield  {author} {\bibinfo {author} {\bibfnamefont {H.}~\bibnamefont
  {{Sakamoto, H. Haniu}}},\ }\href {\doibase 10.1115/1.2909415} {\bibfield
  {journal} {\bibinfo  {journal} {Journal of Fluid Engineering}\ }\textbf
  {\bibinfo {volume} {112}},\ \bibinfo {pages} {386} (\bibinfo {year}
  {1990})}\BibitemShut {NoStop}%
\bibitem [{\citenamefont {Mittal}(1999)}]{Mittal1999}%
  \BibitemOpen
  \bibfield  {author} {\bibinfo {author} {\bibfnamefont {R.}~\bibnamefont
  {Mittal}},\ }\href {\doibase 10.2514/3.14179} {\bibfield  {journal} {\bibinfo
   {journal} {AIAA Journal}\ }\textbf {\bibinfo {volume} {37}},\ \bibinfo
  {pages} {388} (\bibinfo {year} {1999})}\BibitemShut {NoStop}%
\bibitem [{\citenamefont {Tomboulides}\ and\ \citenamefont
  {Orszag}(2000)}]{Tomboulides2000}%
  \BibitemOpen
  \bibfield  {author} {\bibinfo {author} {\bibfnamefont {A.}~\bibnamefont
  {Tomboulides}}\ and\ \bibinfo {author} {\bibfnamefont {S.}~\bibnamefont
  {Orszag}},\ }\href@noop {} {\bibfield  {journal} {\bibinfo  {journal}
  {Journal of Fluid Mechanics}\ }\textbf {\bibinfo {volume} {416}},\ \bibinfo
  {pages} {45} (\bibinfo {year} {2000})}\BibitemShut {NoStop}%
\bibitem [{\citenamefont {Constantinescu}\ and\ \citenamefont
  {Squires}(2003)}]{Constantinescu2003}%
  \BibitemOpen
  \bibfield  {author} {\bibinfo {author} {\bibfnamefont {G.~S.}\ \bibnamefont
  {Constantinescu}}\ and\ \bibinfo {author} {\bibfnamefont {K.~D.}\
  \bibnamefont {Squires}},\ }\href {\doibase
  10.1023/B:APPL.0000004937.34078.71} {\bibfield  {journal} {\bibinfo
  {journal} {Flow, Turbulence and Combustion}\ }\textbf {\bibinfo {volume}
  {70}},\ \bibinfo {pages} {267} (\bibinfo {year} {2003})}\BibitemShut
  {NoStop}%
\bibitem [{\citenamefont {Constantinescu}\ and\ \citenamefont
  {Squires}(2004)}]{Constantinescu2004}%
  \BibitemOpen
  \bibfield  {author} {\bibinfo {author} {\bibfnamefont {G.}~\bibnamefont
  {Constantinescu}}\ and\ \bibinfo {author} {\bibfnamefont {K.}~\bibnamefont
  {Squires}},\ }\href {\doibase 10.1063/1.1688325} {\bibfield  {journal}
  {\bibinfo  {journal} {Physics of Fluids}\ }\textbf {\bibinfo {volume} {16}},\
  \bibinfo {pages} {1449} (\bibinfo {year} {2004})}\BibitemShut {NoStop}%
\bibitem [{\citenamefont {Yun}\ \emph {et~al.}(2006)\citenamefont {Yun},
  \citenamefont {Kim},\ and\ \citenamefont {Choi}}]{Yun2006}%
  \BibitemOpen
  \bibfield  {author} {\bibinfo {author} {\bibfnamefont {G.}~\bibnamefont
  {Yun}}, \bibinfo {author} {\bibfnamefont {D.}~\bibnamefont {Kim}}, \ and\
  \bibinfo {author} {\bibfnamefont {H.}~\bibnamefont {Choi}},\ }\href {\doibase
  10.1063/1.2166454} {\bibfield  {journal} {\bibinfo  {journal} {Physics of
  Fluids}\ ,\ \bibinfo {pages} {015102}} (\bibinfo {year} {2006})}\BibitemShut
  {NoStop}%
\bibitem [{\citenamefont {Hoffman}(2006)}]{hoffman2006}%
  \BibitemOpen
  \bibfield  {author} {\bibinfo {author} {\bibfnamefont {J.}~\bibnamefont
  {Hoffman}},\ }\href {\doibase 10.1017/S0022112006002679} {\bibfield
  {journal} {\bibinfo  {journal} {Journal of Fluid Mechanics}\ }\textbf
  {\bibinfo {volume} {568}},\ \bibinfo {pages} {77} (\bibinfo {year}
  {2006})}\BibitemShut {NoStop}%
\bibitem [{\citenamefont {Jones}\ and\ \citenamefont
  {Clarke}(2008)}]{Jones2008}%
  \BibitemOpen
  \bibfield  {author} {\bibinfo {author} {\bibfnamefont {D.~A.}\ \bibnamefont
  {Jones}}\ and\ \bibinfo {author} {\bibfnamefont {D.~B.}\ \bibnamefont
  {Clarke}},\ }\href@noop {} {\emph {\bibinfo {title} {{Simulation of flow past
  a sphere using the Fluent code}}}},\ \bibinfo {type} {Tech. Rep.}\ (\bibinfo
  {institution} {Defence Science and Technology Organisation},\ \bibinfo {year}
  {2008})\BibitemShut {NoStop}%
\bibitem [{\citenamefont {Hadamard}(1911)}]{Hadamard11}%
  \BibitemOpen
  \bibfield  {author} {\bibinfo {author} {\bibfnamefont {J.~S.}\ \bibnamefont
  {Hadamard}},\ }\href@noop {} {\bibfield  {journal} {\bibinfo  {journal} {C.
  R. Acad. Sci. Paris}\ }\textbf {\bibinfo {volume} {152}},\ \bibinfo {pages}
  {1735?1738} (\bibinfo {year} {1911})}\BibitemShut {NoStop}%
\bibitem [{\citenamefont {Rybczynski}(1911)}]{Rybczynski11}%
  \BibitemOpen
  \bibfield  {author} {\bibinfo {author} {\bibfnamefont {W.}~\bibnamefont
  {Rybczynski}},\ }\href@noop {} {\bibfield  {journal} {\bibinfo  {journal}
  {Bull. Acad. Sci. Cracovie A}\ ,\ \bibinfo {pages} {40}} (\bibinfo {year}
  {1911})}\BibitemShut {NoStop}%
\bibitem [{\citenamefont {Moore}(1963)}]{Moore1963}%
  \BibitemOpen
  \bibfield  {author} {\bibinfo {author} {\bibfnamefont {D.~W.}\ \bibnamefont
  {Moore}},\ }\href {\doibase 10.1017/S0022112063000665} {\bibfield  {journal}
  {\bibinfo  {journal} {Journal of Fluid Mechanics}\ }\textbf {\bibinfo
  {volume} {16}},\ \bibinfo {pages} {161} (\bibinfo {year} {1963})}\BibitemShut
  {NoStop}%
\bibitem [{\citenamefont {Shirtcliffe}\ \emph {et~al.}(2009)\citenamefont
  {Shirtcliffe}, \citenamefont {McHale}, \citenamefont {Newton},\ and\
  \citenamefont {Zhang}}]{Shirtcliffe2009}%
  \BibitemOpen
  \bibfield  {author} {\bibinfo {author} {\bibfnamefont {N.~J.}\ \bibnamefont
  {Shirtcliffe}}, \bibinfo {author} {\bibfnamefont {G.}~\bibnamefont {McHale}},
  \bibinfo {author} {\bibfnamefont {M.~I.}\ \bibnamefont {Newton}}, \ and\
  \bibinfo {author} {\bibfnamefont {Y.}~\bibnamefont {Zhang}},\ }\href@noop {}
  {\bibfield  {journal} {\bibinfo  {journal} {ACS applied materials \&
  interfaces}\ }\textbf {\bibinfo {volume} {1}},\ \bibinfo {pages} {1316}
  (\bibinfo {year} {2009})}\BibitemShut {NoStop}%
\bibitem [{\citenamefont {McHale}\ \emph {et~al.}(2009)\citenamefont {McHale},
  \citenamefont {Shirtcliffe}, \citenamefont {Evans},\ and\ \citenamefont
  {Newton}}]{McHale2009}%
  \BibitemOpen
  \bibfield  {author} {\bibinfo {author} {\bibfnamefont {G.}~\bibnamefont
  {McHale}}, \bibinfo {author} {\bibfnamefont {N.~J.}\ \bibnamefont
  {Shirtcliffe}}, \bibinfo {author} {\bibfnamefont {C.~R.}\ \bibnamefont
  {Evans}}, \ and\ \bibinfo {author} {\bibfnamefont {M.~I.}\ \bibnamefont
  {Newton}},\ }\href@noop {} {\bibfield  {journal} {\bibinfo  {journal}
  {Applied Physics Letters}\ }\textbf {\bibinfo {volume} {94}},\ \bibinfo
  {pages} {064104} (\bibinfo {year} {2009})}\BibitemShut {NoStop}%
\bibitem [{\citenamefont {McHale}\ \emph {et~al.}(2010)\citenamefont {McHale},
  \citenamefont {Newton},\ and\ \citenamefont {Shirtcliffe}}]{McHale2010}%
  \BibitemOpen
  \bibfield  {author} {\bibinfo {author} {\bibfnamefont {G.}~\bibnamefont
  {McHale}}, \bibinfo {author} {\bibfnamefont {M.~I.}\ \bibnamefont {Newton}},
  \ and\ \bibinfo {author} {\bibfnamefont {N.~J.}\ \bibnamefont
  {Shirtcliffe}},\ }\href {\doibase 10.1039/b917861a} {\bibfield  {journal}
  {\bibinfo  {journal} {Soft Matter}\ }\textbf {\bibinfo {volume} {6}},\
  \bibinfo {pages} {714} (\bibinfo {year} {2010})}\BibitemShut {NoStop}%
\bibitem [{\citenamefont {McHale}\ \emph {et~al.}(2011)\citenamefont {McHale},
  \citenamefont {Flynn},\ and\ \citenamefont {Newton}}]{McHale2011}%
  \BibitemOpen
  \bibfield  {author} {\bibinfo {author} {\bibfnamefont {G.}~\bibnamefont
  {McHale}}, \bibinfo {author} {\bibfnamefont {M.~R.}\ \bibnamefont {Flynn}}, \
  and\ \bibinfo {author} {\bibfnamefont {M.~I.}\ \bibnamefont {Newton}},\
  }\href {\doibase 10.1039/c1sm06140b} {\bibfield  {journal} {\bibinfo
  {journal} {Soft Matter}\ }\textbf {\bibinfo {volume} {7}},\ \bibinfo {pages}
  {10100} (\bibinfo {year} {2011})}\BibitemShut {NoStop}%
\bibitem [{\citenamefont {Vakarelski}\ \emph {et~al.}(2011)\citenamefont
  {Vakarelski}, \citenamefont {Marston}, \citenamefont {Chan},\ and\
  \citenamefont {Thoroddsen}}]{Vakarelski2011}%
  \BibitemOpen
  \bibfield  {author} {\bibinfo {author} {\bibfnamefont {I.~U.}\ \bibnamefont
  {Vakarelski}}, \bibinfo {author} {\bibfnamefont {J.~O.}\ \bibnamefont
  {Marston}}, \bibinfo {author} {\bibfnamefont {D.~Y.~C.}\ \bibnamefont
  {Chan}}, \ and\ \bibinfo {author} {\bibfnamefont {S.~T.}\ \bibnamefont
  {Thoroddsen}},\ }\href {\doibase 10.1103/PhysRevLett.106.214501} {\bibfield
  {journal} {\bibinfo  {journal} {Physical Review Letters}\ }\textbf {\bibinfo
  {volume} {106}},\ \bibinfo {pages} {3} (\bibinfo {year} {2011})}\BibitemShut
  {NoStop}%
\bibitem [{\citenamefont {Vakarelski}\ \emph {et~al.}(2012)\citenamefont
  {Vakarelski}, \citenamefont {Patankar}, \citenamefont {Marston},
  \citenamefont {Chan},\ and\ \citenamefont {Thoroddsen}}]{Vakarelski2012}%
  \BibitemOpen
  \bibfield  {author} {\bibinfo {author} {\bibfnamefont {I.~U.}\ \bibnamefont
  {Vakarelski}}, \bibinfo {author} {\bibfnamefont {N.~A.}\ \bibnamefont
  {Patankar}}, \bibinfo {author} {\bibfnamefont {J.~O.}\ \bibnamefont
  {Marston}}, \bibinfo {author} {\bibfnamefont {D.~Y.~C.}\ \bibnamefont
  {Chan}}, \ and\ \bibinfo {author} {\bibfnamefont {S.~T.}\ \bibnamefont
  {Thoroddsen}},\ }\href {\doibase 10.1038/nature11418} {\bibfield  {journal}
  {\bibinfo  {journal} {Nature}\ }\textbf {\bibinfo {volume} {489}},\ \bibinfo
  {pages} {274} (\bibinfo {year} {2012})}\BibitemShut {NoStop}%
\bibitem [{\citenamefont {Vakarelski}\ \emph {et~al.}(2013)\citenamefont
  {Vakarelski}, \citenamefont {Chan}, \citenamefont {Marston},\ and\
  \citenamefont {Thoroddsen}}]{Vakarelski2013}%
  \BibitemOpen
  \bibfield  {author} {\bibinfo {author} {\bibfnamefont {I.~U.}\ \bibnamefont
  {Vakarelski}}, \bibinfo {author} {\bibfnamefont {D.~Y.~C.}\ \bibnamefont
  {Chan}}, \bibinfo {author} {\bibfnamefont {J.~O.}\ \bibnamefont {Marston}}, \
  and\ \bibinfo {author} {\bibfnamefont {S.~T.}\ \bibnamefont {Thoroddsen}},\
  }\href {\doibase 10.1021/la402306c} {\bibfield  {journal} {\bibinfo
  {journal} {Langmuir}\ }\textbf {\bibinfo {volume} {29}},\ \bibinfo {pages}
  {11074} (\bibinfo {year} {2013})}\BibitemShut {NoStop}%
\bibitem [{\citenamefont {Vakarelski}\ \emph {et~al.}(2014)\citenamefont
  {Vakarelski}, \citenamefont {Chan},\ and\ \citenamefont
  {Thoroddsen}}]{Vakarelski2014}%
  \BibitemOpen
  \bibfield  {author} {\bibinfo {author} {\bibfnamefont {I.~U.}\ \bibnamefont
  {Vakarelski}}, \bibinfo {author} {\bibfnamefont {D.~Y.~C.}\ \bibnamefont
  {Chan}}, \ and\ \bibinfo {author} {\bibfnamefont {S.~T.}\ \bibnamefont
  {Thoroddsen}},\ }\href {\doibase 10.1039/c4sm00368c} {\bibfield  {journal}
  {\bibinfo  {journal} {Soft Matter}\ }\textbf {\bibinfo {volume} {10}},\
  \bibinfo {pages} {5662} (\bibinfo {year} {2014})}\BibitemShut {NoStop}%
\bibitem [{\citenamefont {Vakarelski}\ \emph {et~al.}(2015)\citenamefont
  {Vakarelski}, \citenamefont {Chan},\ and\ \citenamefont
  {Thoroddsen}}]{Vakarelski2015}%
  \BibitemOpen
  \bibfield  {author} {\bibinfo {author} {\bibfnamefont {I.~U.}\ \bibnamefont
  {Vakarelski}}, \bibinfo {author} {\bibfnamefont {D.~Y.~C.}\ \bibnamefont
  {Chan}}, \ and\ \bibinfo {author} {\bibfnamefont {S.~T.}\ \bibnamefont
  {Thoroddsen}},\ }\href {\doibase 10.1103/PhysRevLett.115.044501} {\bibfield
  {journal} {\bibinfo  {journal} {Physical Review Letters}\ }\textbf {\bibinfo
  {volume} {044501}},\ \bibinfo {pages} {1} (\bibinfo {year}
  {2015})}\BibitemShut {NoStop}%
\bibitem [{\citenamefont {Vakarelski}\ \emph {et~al.}(2016)\citenamefont
  {Vakarelski}, \citenamefont {Berry}, \citenamefont {Chan},\ and\
  \citenamefont {Thoroddsen}}]{Vakarelski2016}%
  \BibitemOpen
  \bibfield  {author} {\bibinfo {author} {\bibfnamefont {I.~U.}\ \bibnamefont
  {Vakarelski}}, \bibinfo {author} {\bibfnamefont {J.~D.}\ \bibnamefont
  {Berry}}, \bibinfo {author} {\bibfnamefont {D.~Y.}\ \bibnamefont {Chan}}, \
  and\ \bibinfo {author} {\bibfnamefont {S.~T.}\ \bibnamefont {Thoroddsen}},\
  }\href {\doibase 10.1103/PhysRevLett.117.114503} {\bibfield  {journal}
  {\bibinfo  {journal} {Physical Review Letters}\ }\textbf {\bibinfo {volume}
  {117}},\ \bibinfo {pages} {114503} (\bibinfo {year} {2016})}\BibitemShut
  {NoStop}%
\bibitem [{\citenamefont {Clift}\ \emph {et~al.}(1978)\citenamefont {Clift},
  \citenamefont {Grace},\ and\ \citenamefont {Weber}}]{Clift1978}%
  \BibitemOpen
  \bibfield  {author} {\bibinfo {author} {\bibfnamefont {R.}~\bibnamefont
  {Clift}}, \bibinfo {author} {\bibfnamefont {J.}~\bibnamefont {Grace}}, \ and\
  \bibinfo {author} {\bibfnamefont {M.}~\bibnamefont {Weber}},\ }\href@noop {}
  {\emph {\bibinfo {title} {{Bubbles, drops, and particles}}}}\ (\bibinfo
  {publisher} {Academic Press},\ \bibinfo {address} {New York},\ \bibinfo
  {year} {1978})\BibitemShut {NoStop}%
\bibitem [{\citenamefont {Schlichting}()}]{Schlichting1979}%
  \BibitemOpen
  \bibfield  {author} {\bibinfo {author} {\bibfnamefont {H.}~\bibnamefont
  {Schlichting}},\ }\href {\doibase 10.1016/S0997-7546(00)01101-8} {\emph
  {\bibinfo {title} {{Boundary-layer theory}}}}\ (\bibinfo  {publisher}
  {McGraw-Hill},\ \bibinfo {address} {New York})\BibitemShut {NoStop}%
\bibitem [{\citenamefont {McHale}(2017)}]{McHale2017}%
  \BibitemOpen
  \bibfield  {author} {\bibinfo {author} {\bibfnamefont {G.}~\bibnamefont
  {McHale}},\ }in\ \href {\doibase 10.1039/9781782621546-00253} {\emph
  {\bibinfo {booktitle} {Non-wettable Surfaces: Theory, Preparation, and
  Applications}}},\ \bibinfo {editor} {edited by\ \bibinfo {editor}
  {\bibfnamefont {R.~H.~A.}\ \bibnamefont {Ras}}\ and\ \bibinfo {editor}
  {\bibfnamefont {A.}~\bibnamefont {Marmur}}}\ (\bibinfo  {publisher} {The
  Royal Society of Chemistry},\ \bibinfo {address} {Cambridge},\ \bibinfo
  {year} {2017})\ pp.\ \bibinfo {pages} {253--284}\BibitemShut {NoStop}%
\bibitem [{\citenamefont {Gruncell}\ \emph {et~al.}(2013)\citenamefont
  {Gruncell}, \citenamefont {Sandham},\ and\ \citenamefont
  {McHale}}]{Gruncell2013}%
  \BibitemOpen
  \bibfield  {author} {\bibinfo {author} {\bibfnamefont {B.~R.}\ \bibnamefont
  {Gruncell}}, \bibinfo {author} {\bibfnamefont {N.~D.}\ \bibnamefont
  {Sandham}}, \ and\ \bibinfo {author} {\bibfnamefont {G.}~\bibnamefont
  {McHale}},\ }\href@noop {} {\bibfield  {journal} {\bibinfo  {journal}
  {Physics of Fluids (1994-present)}\ }\textbf {\bibinfo {volume} {25}},\
  \bibinfo {pages} {043601} (\bibinfo {year} {2013})}\BibitemShut {NoStop}%
\bibitem [{\citenamefont {Lauga}\ and\ \citenamefont
  {Stone}(2003)}]{Lauga2003}%
  \BibitemOpen
  \bibfield  {author} {\bibinfo {author} {\bibfnamefont {E.}~\bibnamefont
  {Lauga}}\ and\ \bibinfo {author} {\bibfnamefont {H.}~\bibnamefont {Stone}},\
  }\href {\doibase 10.1017/S0022112003004695} {\bibfield  {journal} {\bibinfo
  {journal} {J. Fluid Mech}\ }\textbf {\bibinfo {volume} {489}},\ \bibinfo
  {pages} {55} (\bibinfo {year} {2003})}\BibitemShut {NoStop}%
\bibitem [{\citenamefont {Lauga}\ \emph {et~al.}(2007)\citenamefont {Lauga},
  \citenamefont {Brenner},\ and\ \citenamefont {Stone}}]{Lauga2007}%
  \BibitemOpen
  \bibfield  {author} {\bibinfo {author} {\bibfnamefont {E.}~\bibnamefont
  {Lauga}}, \bibinfo {author} {\bibfnamefont {M.~P.}\ \bibnamefont {Brenner}},
  \ and\ \bibinfo {author} {\bibfnamefont {H.~A.}\ \bibnamefont {Stone}},\ }in\
  \href@noop {} {\emph {\bibinfo {booktitle} {Springer Handbook of Experimental
  Fluid Mechanics}}}\ (\bibinfo  {publisher} {Springer},\ \bibinfo {address}
  {Berlin},\ \bibinfo {year} {2007})\ pp.\ \bibinfo {pages} {1219--1240},\
  \Eprint {http://arxiv.org/abs/0501557v3} {arXiv:0501557v3 [arXiv:cond-mat]}
  \BibitemShut {NoStop}%
\bibitem [{\citenamefont {Vinogradova}(1995)}]{Vinogradova1995}%
  \BibitemOpen
  \bibfield  {author} {\bibinfo {author} {\bibfnamefont {O.~I.}\ \bibnamefont
  {Vinogradova}},\ }\href@noop {} {\bibfield  {journal} {\bibinfo  {journal}
  {Langmuir}\ }\textbf {\bibinfo {volume} {11}},\ \bibinfo {pages} {2213}
  (\bibinfo {year} {1995})}\BibitemShut {NoStop}%
\bibitem [{\citenamefont {Tretheway}\ and\ \citenamefont
  {Meinhart}(2004)}]{Tretheway2004}%
  \BibitemOpen
  \bibfield  {author} {\bibinfo {author} {\bibfnamefont {D.~C.}\ \bibnamefont
  {Tretheway}}\ and\ \bibinfo {author} {\bibfnamefont {C.~D.}\ \bibnamefont
  {Meinhart}},\ }\href {\doibase http://dx.doi.org/10.1063/1.1669400}
  {\bibfield  {journal} {\bibinfo  {journal} {Physics of Fluids}\ }\textbf
  {\bibinfo {volume} {16}},\ \bibinfo {pages} {1509} (\bibinfo {year}
  {2004})}\BibitemShut {NoStop}%
\bibitem [{\citenamefont {Busse}\ \emph {et~al.}(2013)\citenamefont {Busse},
  \citenamefont {Sandham}, \citenamefont {McHale},\ and\ \citenamefont
  {Newton}}]{Busse2013}%
  \BibitemOpen
  \bibfield  {author} {\bibinfo {author} {\bibfnamefont {A.}~\bibnamefont
  {Busse}}, \bibinfo {author} {\bibfnamefont {N.~D.}\ \bibnamefont {Sandham}},
  \bibinfo {author} {\bibfnamefont {G.}~\bibnamefont {McHale}}, \ and\ \bibinfo
  {author} {\bibfnamefont {M.~I.}\ \bibnamefont {Newton}},\ }\href {\doibase
  10.1017/jfm.2013.284} {\bibfield  {journal} {\bibinfo  {journal} {Journal of
  Fluid Mechanics}\ }\textbf {\bibinfo {volume} {727}},\ \bibinfo {pages} {488}
  (\bibinfo {year} {2013})}\BibitemShut {NoStop}%
\bibitem [{\citenamefont {Saranadhi}\ \emph {et~al.}(2016)\citenamefont
  {Saranadhi}, \citenamefont {Chen}, \citenamefont {Kleingartner},
  \citenamefont {Srinivasan}, \citenamefont {Cohen},\ and\ \citenamefont
  {McKinley}}]{Saranadhi2016}%
  \BibitemOpen
  \bibfield  {author} {\bibinfo {author} {\bibfnamefont {D.}~\bibnamefont
  {Saranadhi}}, \bibinfo {author} {\bibfnamefont {D.}~\bibnamefont {Chen}},
  \bibinfo {author} {\bibfnamefont {J.~A.}\ \bibnamefont {Kleingartner}},
  \bibinfo {author} {\bibfnamefont {S.}~\bibnamefont {Srinivasan}}, \bibinfo
  {author} {\bibfnamefont {R.~E.}\ \bibnamefont {Cohen}}, \ and\ \bibinfo
  {author} {\bibfnamefont {G.~H.}\ \bibnamefont {McKinley}},\ }\href {\doibase
  10.1126/sciadv.1600686} {\bibfield  {journal} {\bibinfo  {journal} {Science
  Advances}\ }\textbf {\bibinfo {volume} {2}},\ \bibinfo {pages} {e1600686}
  (\bibinfo {year} {2016})}\BibitemShut {NoStop}%
\bibitem [{\citenamefont {Lamb}(1932)}]{Lamb1932}%
  \BibitemOpen
  \bibfield  {author} {\bibinfo {author} {\bibfnamefont {H.}~\bibnamefont
  {Lamb}},\ }\href@noop {} {\emph {\bibinfo {title} {{Hydrodynamics}}}}\
  (\bibinfo  {publisher} {Dover Publications},\ \bibinfo {address} {New York},\
  \bibinfo {year} {1932})\BibitemShut {NoStop}%
\bibitem [{\citenamefont {Swan}\ and\ \citenamefont {Khair}(2008)}]{Swan2008}%
  \BibitemOpen
  \bibfield  {author} {\bibinfo {author} {\bibfnamefont {J.~W.}\ \bibnamefont
  {Swan}}\ and\ \bibinfo {author} {\bibfnamefont {A.~S.}\ \bibnamefont
  {Khair}},\ }\href {\doibase 10.1017/S0022112008001614} {\bibfield  {journal}
  {\bibinfo  {journal} {Journal of Fluid Mechanics}\ }\textbf {\bibinfo
  {volume} {606}},\ \bibinfo {pages} {115} (\bibinfo {year}
  {2008})}\BibitemShut {NoStop}%
\bibitem [{\citenamefont {Germano}\ \emph {et~al.}(1991)\citenamefont
  {Germano}, \citenamefont {Piomelli}, \citenamefont {Moin},\ and\
  \citenamefont {Cabot}}]{Germano1993}%
  \BibitemOpen
  \bibfield  {author} {\bibinfo {author} {\bibfnamefont {M.}~\bibnamefont
  {Germano}}, \bibinfo {author} {\bibfnamefont {U.}~\bibnamefont {Piomelli}},
  \bibinfo {author} {\bibfnamefont {P.}~\bibnamefont {Moin}}, \ and\ \bibinfo
  {author} {\bibfnamefont {W.~H.}\ \bibnamefont {Cabot}},\ }\href@noop {}
  {\bibfield  {journal} {\bibinfo  {journal} {Physics of Fluids}\ ,\ \bibinfo
  {pages} {1760}} (\bibinfo {year} {1991})}\BibitemShut {NoStop}%
\bibitem [{\citenamefont {Liovic}\ and\ \citenamefont
  {Lakehal}(2007)}]{Liovic2007}%
  \BibitemOpen
  \bibfield  {author} {\bibinfo {author} {\bibfnamefont {P.}~\bibnamefont
  {Liovic}}\ and\ \bibinfo {author} {\bibfnamefont {D.}~\bibnamefont
  {Lakehal}},\ }\href@noop {} {\bibfield  {journal} {\bibinfo  {journal}
  {Journal of Computational Physics}\ }\textbf {\bibinfo {volume} {222}},\
  \bibinfo {pages} {504} (\bibinfo {year} {2007})}\BibitemShut {NoStop}%
\end{thebibliography}
\end{document}